\documentclass[12pt]{article}

\usepackage[utf8]{inputenc}

\usepackage{url}
\usepackage{amsmath}
\usepackage{amssymb}
\usepackage[round]{natbib}
\usepackage{siunitx}
\usepackage{setspace}
\usepackage{mathtools}
\usepackage{graphicx}
\usepackage{color}
\usepackage{enumerate}
\usepackage{algorithm}
\usepackage{algpseudocodex}
\usepackage{ifthen}
\usepackage{textcomp} 
\usepackage[labelformat=simple]{subcaption}

\usepackage{fancyvrb}

\definecolor{violet}{rgb}{0.7,0,0.7}
\definecolor{gray}{rgb}{0.6,0.6,0.6}

\usetikzlibrary{positioning, arrows.meta, shapes.geometric, patterns}
\tikzset{%
	thick/.style = {line width = 1pt},
	-{Latex},
	>={Latex},
	every edge/.append style={semithick},
	obs/.style = {name = #1, circle, draw, inner sep = 6pt, label = center:$#1$},
	  sb/.style n args = {4}{name=#3, circle, draw, inner sep=1pt, minimum size=7pt, label={[shift={(#1,#2)}]#4:$#3$}},
	clu/.style = {name = #1, regular polygon, regular polygon sides = 4, draw, fill = gray, inner sep = 4pt, label = center:$\+{#1}$}
}

\newcommand{\+}[1]{\ensuremath{\mathbf{#1}}}

\newcommand{\cond}{\,\vert\,}
\newcommand{\given}{{ \, | \, }}

\newcommand{\Pa}[1][]{%
  \ifthenelse{ \equal{#1}{} }
    {\textrm{Pa}}
    {\textrm{Pa}_{#1}}
}
\newcommand{\Ch}[1][]{%
  \ifthenelse{ \equal{#1}{} }
    {\textrm{Ch}}
    {\textrm{Ch}_{#1}}
}
\newcommand{\An}[1][]{%
  \ifthenelse{ \equal{#1}{} }
    {\textrm{An}}
    {\textrm{An}_{#1}}
}
\newcommand{\De}[1][]{%
  \ifthenelse{ \equal{#1}{} }
    {\textrm{De}}
    {\textrm{De}_{#1}}
}
\newcommand{\Ne}[1][]{%
  \ifthenelse{ \equal{#1}{} }
    {\textrm{Ne}}
    {\textrm{Ne}_{#1}}
}
\newcommand{\Co}[1][]{%
  \ifthenelse{ \equal{#1}{} }
    {\textrm{Co}}
    {\textrm{Co}_{#1}}
}
\newcommand{\rec}[1][]{%
  \ifthenelse{ \equal{#1}{} }
    {\textrm{Re}}
    {\textrm{Re}_{#1}}
}
\newcommand{\emi}[1][]{%
  \ifthenelse{ \equal{#1}{} }
    {\textrm{Em}}
    {\textrm{Em}_{#1}}
}

\AtBeginDocument{%
  \DeclareFontShape{OT1}{cmr}{m}{scit}{<->ssub*cmr/m/sc}{}%
}

\newcommand{\eref}[1]{(\ref{#1})}

\newcommand{\doo}{\textrm{do}}

\definecolor{colA}{RGB}{241,86,63} 
\definecolor{colB}{RGB}{0,82,174} 
\definecolor{colC}{RGB}{129,103,0}

\newtheorem{definition}{Definition}

\title{Generalizing experimental findings: identification beyond adjustments}
\author{Juha Karvanen\\
Department of Mathematics and Statistics,\\ University of Jyvaskyla, Finland\\
juha.t.karvanen@jyu.fi}
\date{February 14, 2022}

\begin{document}

\maketitle

\begin{abstract} 
{We aim to generalize the results of a randomized controlled trial (RCT) to a target population with the help of some observational data. This is a problem of causal effect identification with multiple data sources. Challenges arise when the RCT is conducted in a context that differs from the target population. Earlier research has focused on cases where the estimates from the RCT can be adjusted by observational data in order to remove the selection bias and other domain specific differences.
We consider examples where the experimental findings cannot be generalized by an adjustment and show that the generalization may still be possible by other identification strategies that can be derived by applying do-calculus. The obtained identifying functionals for these examples contain trapdoor variables of a new type. The value of a trapdoor variable needs to be fixed in the estimation and the choice of the value may have a major effect on the bias and accuracy of estimates, which is also seen in simulations. 
The presented results expand the scope of settings where the generalization of experimental findings is doable.
}
\end{abstract}
%
%


\section{Introduction}
The generalization of experimental findings from a population to another requires careful consideration of the potential sources of bias and understanding about the differences between the populations. These differences and biases can be described and analyzed using the tools of causal inference \citep{pearl2014external,pearl2015generalizing}. The goal is then to identify the causal effect of interest in the target population using data sources that originate from other populations.

We consider settings where available information comes from a randomized controlled trial (RCT) and one or more surveys. Under ideal conditions, i.e, without missing data, non-compliance, measurement error, etc., the results from an RCT estimate causal effects in the domain where the trial has been conducted. Due to costs and practical difficulties, the sample sizes in RCTs are typically small or moderate. 

Differently from ideal RCTs, unobserved confounders may prevent the identification of a causal effect in a survey. In surveys, the sample sizes are typically larger than in RCTs but variables that require, for instance, laboratory assessments or physical measurements may not be available. This means that the variables in the RCT and the survey may be partially non-overlapping, which causes an additional challenge for causal inference.

The process of causal inference consists of three phases: modeling, identification and estimation.
First, the nonparametric causal model for the phenomenon of interest is formulated using directed acyclic graphs and their extensions. The graph can be amended by elements that describe the study design, selection bias and missing data \citep{pearl2014external,karvanen2015study,thoemmes2015graphical}. A special notation can be also used to indicate the differences between the populations. 

In the next phase, algorithmic tools are applied to find out whether the causal effect of interest is identifiable from the available data sources under the specified nonparameteric causal model. These tools include the ID algorithm \citep{shpitser2006id} and its extensions to transportability \citep{bareinboim2014transportability} and selections bias \citep{bareinboim2015recovering}. Software implementations are available in R \citep{tikka2017b}. The most general tool for identification problems is do-calculus for which a search-based algorithmic implementation also exists \citep{dosearchR,tikka2021dosearch}.

The final phase includes the estimation of causal effect according to the formula obtained with an identification algorithm. A special challenge occurs when a trapdoor variable \citep{helske2021estimation} is present in the identifying formula. A trapdoor variable is a consequence of functional equality constraints known as Verma-constraints
 \citep{robins1986,TianPearl2002,pearl1991} and in the identifying formula it must be fixed to some value. The choice of this value may affect the precision of estimation, there may be a considerable bias when the sample size is small \citep{helske2021estimation}.

Generalizability, transportability and external validity have been active research topics in recent years, e.g. \citep{lesko2017generalizing,buchanan2018generalizing,westreich2019target, dahabreh2019generalizing, dahabreh2020toward}, see also reviews \citep{colnet2020causal,degtiar2021review}.
Earlier works 
focusing on identification seem to have concentrated on covariate adjustments \citep{pearl:book2009} as a way to standardize the experimental results to the target population \citep{pearl2014external,correa2019adjustment}. An adjustment is formally defined in Section~\ref{sec:notation} and refers here to a summation formula similar to the well-known back-door adjustment. \citet{correa2019adjustment} have proposed an algorithm for enumerating all admissible sets for adjustment.

While adjustments are probably the most commonly used identification strategies, there exist also other strategies, such as the front-door criterion \citep{pearl1995} and  more complicated identifying functionals \citep{tikka2017b,tikka2021dosearch}, instrumental variables \citep{angrist1996identification}, proxy variables \citep{miao2018identifying}, and context-specific independence relations \citep{tikka2020csi}. These identification strategies are important because they sometimes work even if no adjustments are possible.

In this paper, we present strategies for generalizing experimental findings in situations where simple adjustment criteria do not lead to identification. These strategies can be derived using do-calculus \citep{pearl1995} and its algorithmic implementation, \texttt{do-search} \citep{tikka2021dosearch}. Remarkably, the proposed strategies work in examples presented by \citet{correa2019adjustment} as cases where no adjustments are applicable.  
A distinct benefit of \texttt{do-search} is its ability to work with multiple data sources with partially non-overlapping variables. 

As the second contribution, we introduce a new type of trapdoor variable that may occur in data fusion. These trapdoor variables are present in our examples and they are children of the intervened variable. In earlier examples \citep{helske2021estimation}, the trapdoor variable has always been an ancestor of the intervened variable. The presented examples demonstrate that also these new trapdoor variables may induce the small-sample bias.

The rest of the paper is organized as follows. The key concepts are defined in Section~\ref{sec:notation}. Section~\ref{sec:strategies} presents examples that demonstrate identification strategies beyond adjustments.  Section~\ref{sec:numerical} studies the numerical estimation with trapdoor variables in scenarios with different sample sizes. Section~\ref{sec:discussion} concludes the paper.


%
%
%
%
%
%
%
%
%
%
%

\section{Notation and related work} \label{sec:notation}

The reader is assumed to be familiar with structural causal models and their graphical presentation \citep{pearl:book2009}. We denote variables by capital letters and their values by small letters. Sets of variables are denoted in bold. The set of all possible value assignments to variable $\+ X$ is denoted by $val(\+ X)$. The do-operator notation $\doo(X=x)$ or $\doo(X)$ denotes an intervention where the value of variable $X$ is set to be $x$. In graphs, unobserved confounders are denoted by bidirected edges.

\subsection{Identifiability, do-calculus and do-search}
A causal effect is identifiable if it can be expressed as a functional of available data sources. Let $P_1,P_2,\ldots,P_k$ denote input distributions that may be observational or experimental. If causal effect $P(\+Y \cond \doo(\+X))$ is identifiable from $P_1,P_2,\ldots,P_k$,  it can be written as $P(\+Y \cond \doo(\+X))=f(P_1,P_2,\ldots,P_k)$ where $f$ is an identifying functional.

Do-calculus \citep{pearl1995} is a methodology for finding identifying functionals by iterative application of symbolic operations to probabilistic expressions. The approach starts from one or more input distributions, that are presented in a symbolic form, such as $P(X,Y,Z)$.
The rules of do-calculus specify requirements for adding and removing variables, exchanging interventions and observations and removing and adding interventions. In addition, standard probability calculus is needed. The goal is to find a sequence of the rules that transform the input distribution(s) into the causal effect interest (for instance, $P(Y \cond \doo(X))$.

The practical challenge of do-calculus is that the order of the rules to be applied cannot be determined in advance. To address this challenge, identification algorithms that work in polynomial-time have been derived for many important identification problems \citep{shpitser2006id,bareinboim2015recovering,bareinboim2014transportability,lee2019general}.
Despite the progress in algorithm development, do-calculus is still the most general tool available for identification problems. \texttt{Do-search} \citep{tikka2021dosearch} is an implementation of do-calculus that is based on extensive search over the rules of do-calculus and probability calculus. \texttt{Do-search} is not applicable to problems with a large number of variables because of high computational cost but works well with graphs of the size that is typically encountered in the literature on causal inference.

\subsection{Adjustments}
One of the simplest forms of the identifying functional is an adjustment \citep{pearl:book2009,correa2019adjustment} defined as follows
\begin{definition}[Adjustment] \label{def:adjustment}
A causal effect $P(\+Y \cond \doo(\+X))$ is identifiable by an adjustment if for some set of observed variables $\+Z$, $\+Z \cap (\+X \cup \+Y)=\emptyset$ it holds
\end{definition}
\begin{equation} \label{eq:adjustment}
P(\+Y \cond \doo(\+X)) = \sum_{\+Z} P(\+Y \cond \+X,\+Z)P(\+Z).
\end{equation}

Sometimes identifying functionals that do not follow Definition~\ref{def:adjustment} are also called ``adjustments''; consider the front-door adjustment as an example. In this paper, however, ``beyond adjustments'' refers to identifying functionals that cannot be presented in the form of equation~\eref{eq:adjustment}.

\subsection{Transportability and selection bias}

Conceptually, a transportability variable $T$ indicates the potential differences between the target population and the experimental domain. Technically, an edge $T \rightarrow Z$ in the causal diagram implies that distributions $P(Z)$ and $P(Z \cond T)$ may differ from each other. If an experiment where $X$ is intervened and $Y$ measured has been conducted in the experimental domain, we specify input distribution as $P(Y \cond X,T)$. The role of $T$ as a transportability variable is apparent if $T$ does not have parents and none of the inputs available gives information on the distribution of $T$. In a causal diagram, the vertex for a transportability variable is drawn as a filled square.

Conditional distribution $P(X \cond S)$,  where $S$  is a selection variable, indicates selective sampling of $X$ with respect to the target population. If there is an edge $Z \rightarrow S$, the selection depends on variable $Z$. In a causal diagram, the vertex for a selection variable is drawn as a double circle. 

We do not use any special definitions for transportability or selection bias in the formal inference. This means that transportability variables and selection variables are mathematically indistinguishable from ordinary variables in the structural causal model and they are handled in do-calculus like all other variables. However, the role of transportability variables and selection variables can be seen when the input distributions are represented in the symbolic form.
The distributions that directly represent the target population are unconditional while other distributions are conditioned on a transportability variable or a selection variable.


\subsection{Trapdoor variables}

We define trapdoor variables following \citet{helske2021estimation}.
The formal definition of trapdoor variables is based on functional independence defined as follows
\begin{definition}[Functional independence]
		Let $\+ X, \+Y, \+ Z$ be disjoint sets of variables and let ${P(\+Y \given \doo(\+ X  = \+ x))}$ be an identifiable causal effect with an identifying functional $g$. If the domain of $g$ is $val(\+ X) \times val(\+ Y) \times val(\+ Z)$, and $g(\+ x, \+ y, \+ z_1) = g(\+ x, \+ y, \+ z_2)$ for all $\+ x \in val(\+ X)$, $\+y \in val(\+ Y)$ and $\+ z_1,\+ z_2 \in val(\+ Z)$, then $g$ is \emph{functionally independent} from $\+ Z$. 
	\end{definition}


In addition, we need to define latent projection  \citep{pearl1991} that removes some vertices in the graph and updates the edges:
\begin{definition}[Latent projection] Let $G$ be a causal graph over a set of vertices $\+ V \cup \+ L$. The \emph{latent projection} $L(G, \+ V)$ is a causal graph over $\+ V$ where for every pair of distinct vertices $Z,W \in \+ V$ it holds that
		\begin{enumerate}[(a)]
			\item $L(G, \+ V)$ contains an edge $Z \longrightarrow W$ if there exists a directed path $Z \longrightarrow \cdots \longrightarrow W$ in $G$ on which every vertex except $Z$ and $W$ is in $\+ L$.
			\item $L(G, \+ V)$ contains an edge $Z \longleftrightarrow W$ if there exists a path from $Z$ to $W$ in $G$ that does not contain the pattern $Z \longrightarrow M \longleftarrow W$ (a collider), on which every vertex except $Z$ and $W$ is in $\+ L$, the first edge of the path has as arrowhead into $W$, and the last edge has an arrowhead into $Z$.
		\end{enumerate}
	\end{definition}

A trapdoor variable is a variable that is functionally independent from the identifying functional but crucial for the identifiability:  
\begin{definition}[Trapdoor variable] Let $G$ be a causal graph over a set of vertices $\+ V$ and let ${P(\+ Y \given \doo(\+ X = \+ x))}$ be identifiable in $G$ from input distributions $P_1,\ldots,P_k$ with identifying functional $g(P_1,\ldots,P_k)$. Set $\+ Z$,  where $\+ Z \subset \+ V \setminus (\+ X \cup \+ Y)$, is a set of trapdoor variables with respect to $g$ in $G$ if  $g$ is functionally independent of $\+ Z$ and  $P(\+ Y \given \doo(\+ X = \+ x))$ is not identifiable from $P_1,\ldots,P_k$ in $L(G, \+ V \setminus \+ Z)$.
	\end{definition}

\begin{figure}[!ht]
		\centering
			\begin{tikzpicture}[scale=0.9]
			\node [obs = {W}] at (-8,0) {$\vphantom{X}$};
			\node [obs = {Y}] at (1,0) {$\vphantom{X}$};
			\node [obs = {X}] at (-2,0) {$\vphantom{X}$};
			\node [obs = {Z}] at (-5,0) {$\vphantom{X}$};
			\path [->] (X) edge (Y);
			\path [->] (Z) edge (X);
			\path [->] (W) edge (Z);
			\path [<->,dashed] (Y) edge [bend right = 45] (W);
			\path [<->,dashed] (W) edge [bend right = 45] (X);
			\end{tikzpicture}
		\caption{A graph where $Z$ is a trapdoor variable when  the causal effect of $X$ on $Y$ is identified from $P(W,Z,X,Y)$.}
		\label{fig:basictrapdoor}
	\end{figure}
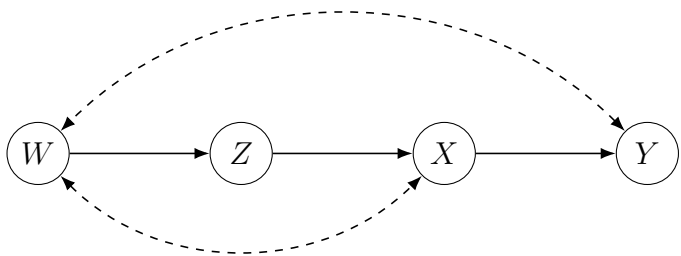

Graph $G$ in Figure~\ref{fig:basictrapdoor} presents a basic example on a trapdoor variable. The causal effect $P(Y \cond \doo(X))$ is identifiable from observations $P(W,Z,X,Y)$ as follows
\begin{align} \label{eq:basictrapdoor}
P(Y \cond \doo(X)) = & g(W,Z,X,Y) = \frac{\sum_{W}P(W)P(X,Y|Z,W)}{\sum_{Y} \sum_{W} P(W)P(X,Y|Z,W)} =  \nonumber \\
& \frac{\sum_{W} P(W)P(X|Z,W)P(Y|X,Z,W)}{ \sum_{W}P(W)P(X|Z,W)}
\end{align}
We notice that value of $Z$ can be freely chosen in equation~\eref{eq:basictrapdoor} and thus $Z$ is functionally independent from $g(W,Z,X,Y)$. Moreover, $P(Y \cond \doo(X))$ is not identifiable in the latent projection $L(G, \{W,X,Y\})$, which makes $Z$ a trapdoor variable. Although the value of a trapdoor variable is arbitrary in theory, \citet{helske2021estimation} demonstrate that a poor choice of the value of a trapdoor variable may lead to considerable bias in estimation with small sample size.

\section{Strategies for generalization} \label{sec:strategies}
We will now show how do-calculus can be used to generalize experimental findings that cannot be generalized by an adjustment (in sense of Definition~\ref{def:adjustment}). We consider two examples that  \citet{correa2019adjustment} present as cases where the causal effect of interest is not identifiable by any adjustment. 

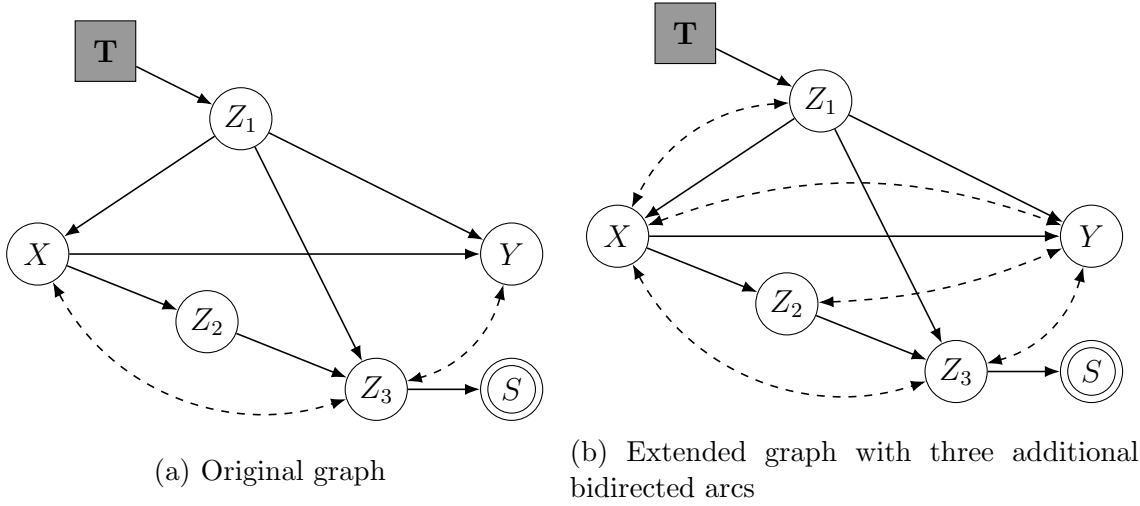
\begin{figure}[!ht]
		\centering
		\begin{subfigure}{0.47\textwidth}
			\centering
			\begin{tikzpicture}[xscale=1,scale = 0.9]
			\node [obs = {X}] at (0,0) {$\vphantom{X}$};			
			\node [obs = {Y}] at (7,0) {$\vphantom{X}$};
			\node [obs = {Z_1}] at (3,2) {$\vphantom{X}$};
			\node [clu = {T}] at (1,3) {$\vphantom{X}$};				
			\node [obs = {Z_2}] at (2.5,-1) {$\vphantom{X}$};			
			\node [obs = {Z_3}] at (5,-2) {$\vphantom{X}$};
			\node [obs = {S}] at (7,-2) {$\vphantom{X}$};	
			\draw [black, inner sep = 0.5pt] (7,-2) circle (10pt);		

			\path [->] (T) edge (Z_1);
			\path [->] (Z_1) edge (X);
			\path [->] (Z_1) edge (Z_3);
			\path [->] (Z_1) edge (Y);
			\path [->] (X) edge (Y);
			\path [->] (X) edge (Z_2);
			\path [->] (Z_2) edge (Z_3);
			\path [->] (Z_3) edge (S);

			\path [<->,dashed] (X) edge [bend right = 40] (Z_3);
			\path [<->,dashed] (Y) edge [bend left = 30] (Z_3);
			\end{tikzpicture}
			\caption{Original graph \label{fig:ex1a}}
		\end{subfigure}
				\begin{subfigure}{0.47\textwidth}
			\centering
			\begin{tikzpicture}[xscale=1,scale = 0.9]
			\node [obs = {X}] at (0,0) {$\vphantom{X}$};			
			\node [obs = {Y}] at (7,0) {$\vphantom{X}$};
			\node [obs = {Z_1}] at (3,2) {$\vphantom{X}$};
			\node [clu = {T}] at (1,3) {$\vphantom{X}$};				
			\node [obs = {Z_2}] at (2.5,-1) {$\vphantom{X}$};			
			\node [obs = {Z_3}] at (5,-2) {$\vphantom{X}$};
			\node [obs = {S}] at (7,-2) {$\vphantom{X}$};	
			\draw [black, inner sep = 0.5pt] (7,-2) circle (10pt);		

			\path [->] (T) edge (Z_1);
			\path [->] (Z_1) edge (X);
			\path [->] (Z_1) edge (Z_3);
			\path [->] (Z_1) edge (Y);
			\path [->] (X) edge (Y);
			\path [->] (X) edge (Z_2);
			\path [->] (Z_2) edge (Z_3);
			\path [->] (Z_3) edge (S);

			\path [<->,dashed] (X) edge [bend right = 40] (Z_3);
			\path [<->,dashed] (Y) edge [bend left = 30] (Z_3);
			\path [<->,dashed] (X) edge [bend left = 20] (Y);			
			\path [<->,dashed] (Z_1) edge [bend right = 30] (X);
			\path [<->,dashed] (Z_2) edge [bend right = 10] (Y);
			\end{tikzpicture}
			\caption{Extended graph with three additional bidirected arcs \label{fig:ex1b}}
		\end{subfigure}
\caption{Example 1: Two graphs where $P(Y \cond \doo(X))$ is identifiable from $P(Y,Z_1,Z_2,Z_3 \cond \doo(X),T,S)$ and $P(Z_1,Z_2,Z_3,X)$. \label{fig:correa1b}} 
\end{figure}

Figure~\ref{fig:ex1a} presents a graph considered by \citet{correa2019adjustment}.  In this example, we are interested in the effect of gene therapy ($X$) on a certain type of leukemia ($Y$). The gene therapy has an adverse effect of decreasing blood cells ($Z_2$), which may lead to anemia and infections ($Z_3$). Background factors ($Z_1$) such as demographics and health history may also have an impact on the adverse effects, leukemia and gene therapy. There are unobserved confounders between the gene therapy and the adverse effects and between the adverse effects and the outcome. 

In an RCT, $X$ is intervened and $Y$, $Z_1$, $Z_2$ and $Z_3$ are measured. The RCT is conducted in domain $T$ that differs from the target population in the terms of the distribution of the background variables $Z_1$. In addition, adverse effects may force some participants to quit the trial. This leads to selection denoted by $S$ in the graph. Note that although variables $Z_2$ and $Z_3$ are not ancestors of the response $Y$ they are essential for the identification because they are on the directed path from $X$ to $S$.

Besides the RCT providing information on $P(Y,Z_1,Z_2,Z_3 \cond \doo(X),T,S)$, we have access to observational data that gives information on the joint distribution $P(Z_1,Z_2,Z_3,X)$ in the target population. 

As noted in \citep{correa2019adjustment}, there is no set that works as an adjustment set according Definition~\ref{def:adjustment}. However, causal effect $P(Y \cond \doo(X))$ is still identifiable from inputs $P(Y \cond \doo(X),Z_1,Z_3,Z_2,T,S)$ and $P(Z_1,Z_2,Z_3,X)$. Application of do-calculus results in
\begin{align} \label{eq:ex1a}
& P(Y \cond \doo(X)) = 
 \sum_{Z_1,Z_3}\left(P(Y \cond \doo(X),Z_1,Z_3,Z_2,T,S)\sum_{X}P(Z_1,X)P(Z_3 \cond Z_1,X,Z_2)\right). 
\end{align}
To see this, we first add $Z_2$ to the intervention and write the expression as a sum over $Z_1$ and $Z_3$
\begin{align*}
P(Y \cond \doo(X)) = P(Y \cond \doo(X,Z_2)) = \sum_{Z_1,Z_3} P(Y \cond \doo(X,Z_2), Z_1,Z_3) P(Z_1,Z_3 \cond  \doo(X,Z_2)).
\end{align*}
We notice that 
\begin{equation*}
P(Y \cond \doo(X,Z_2), Z_1,Z_3) = P(Y \cond \doo(X),Z_1,Z_3,Z_2,T,S)
\end{equation*}
and modify the second term of the sum as follows
\begin{align*}
& P(Z_1,Z_3 \cond  \doo(X,Z_2)) = P(Z_1,Z_3 \cond  \doo(Z_2)) = \\
& \sum_{X} P(Z_3 \cond \doo(Z_2),Z_1,X) P(Z_1,X \cond \doo(Z_2)) = \sum_{X} P(Z_3 \cond Z_2,Z_1,X) P(Z_1,X),
\end{align*}
which leads to equation~\eref{eq:ex1a}.

The outmost summation in equation~\eref{eq:ex1a} goes over variables $Z_1$ and $Z_3$. However, instead of joint distribution $P(Z_1,Z_3)$ that would be present in an adjustment, the formula contains a more complicated term  $\sum_{X}P(Z_1,X)P(Z_3 \cond Z_1,X,Z_2)$ that can be understood as a ``pseudo joint distribution'' of $Z_1$ and $Z_3$. Variable $Z_2$ is a trapdoor variable with respect to equation~\eref{eq:ex1a}  because it is functionally independent from the identifying functional and $P(Y \cond \doo(X))$ is not identifiable in the graph where $Z_2$ is removed by the latent projection. 

Figure~\ref{fig:ex1b} shows an extended version of graph of Figure~\ref{fig:ex1a} with additional unobserved confounders. Even here, causal effect $P(Y \cond \doo(X))$ is identifiable from inputs $P(Y \cond \doo(X),Z_1,Z_3,Z_2,T,S)$ and $P(Z_1,Z_2,Z_3,X)$ as follows
\begin{align} \label{eq:ex1b}
& P(Y \cond \doo(X)) = \nonumber \\
& \sum_{Z_1,Z_3,Z_2}\left(P(Y \cond\doo(X),Z_1,Z_3,Z_2,T,S) P(Z_2 \cond X)\sum_{X}P(Z_1,X)P(Z_3 \cond Z_1,X,Z_2) \right) 
\end{align}
Differently from equation~\ref{eq:ex1a}, equation~\ref{eq:ex1b} includes the term $P(Z_2 \cond X)$ and summation over $Z_2$. This means that $Z_2$ is not a trapdoor variable in equation~\eref{eq:ex1b}.

\begin{figure}[!ht]
		\centering
		\begin{subfigure}{0.47\textwidth}
			\centering
			\begin{tikzpicture}[scale = 1.4]
			\node [obs = {X_1}] at (0,5) {$\vphantom{X}$};	
			\node [obs = {Z_1}] at (0,3.75) {$\vphantom{X}$};	
			\node [obs = {Z_2}] at (0,2.5) {$\vphantom{X}$};	
			\node [obs = {Y}] at (2,3) {$\vphantom{X}$};				
			\node [obs = {S}] at (2,1) {$\vphantom{X}$};	
			\draw [black, inner sep = 0.5pt] (2,1) circle (6pt);	
			\node [obs = {X_2}] at (4,5) {$\vphantom{X}$};	
			\node [obs = {Z_3}] at (4,3.75) {$\vphantom{X}$};	
			\node [obs = {Z_4}] at (4,2.5) {$\vphantom{X}$};				
					
			\path [->] (X_1) edge (Z_1);
			\path [->] (Z_1) edge (Z_2);
			\path [->] (Z_2) edge [bend right = 30]  (S);
			\path [->] (X_1) edge (Y);
			\path [->] (X_2) edge (Z_3);
			\path [->] (Z_3) edge (Z_4);
			\path [->] (Z_4) edge [bend left = 30]  (S);
			\path [->] (X_2) edge (Y);

			\path [<->] (X_1) edge [draw=none, bend left = 30] (X_2);
			\path [<->,dashed] (X_1) edge [bend left = 30] (Z_4);
			\path [<->,dashed] (X_2) edge [bend right = 30] (Z_2);
			\path [<->,dashed] (Y) edge [bend left = 30] (Z_2);
			\path [<->,dashed] (Y) edge [bend right = 30] (Z_4);
			\end{tikzpicture}
			\caption{Original graph \label{fig:correa3aa}}
			\end{subfigure}
					\begin{subfigure}{0.47\textwidth}
			\centering
			\begin{tikzpicture}[scale = 1.4]
			\node [obs = {X_1}] at (0,5) {$\vphantom{X}$};	
			\node [obs = {Z_1}] at (0,3.75) {$\vphantom{X}$};	
			\node [obs = {Z_2}] at (0,2.5) {$\vphantom{X}$};	
			\node [obs = {Y}] at (2,3) {$\vphantom{X}$};				
			\node [obs = {S}] at (2,1) {$\vphantom{X}$};	
			\draw [black, inner sep = 0.5pt] (2,1) circle (6pt);	
			\node [obs = {X_2}] at (4,5) {$\vphantom{X}$};	
			\node [obs = {Z_3}] at (4,3.75) {$\vphantom{X}$};	
			\node [obs = {Z_4}] at (4,2.5) {$\vphantom{X}$};				
					
			\path [->] (X_1) edge (Z_1);
			\path [->] (Z_1) edge (Z_2);
			\path [->] (Z_2) edge [bend right = 30]  (S);
			\path [->] (X_1) edge (Y);
			\path [->] (X_2) edge (Z_3);
			\path [->] (Z_3) edge (Z_4);
			\path [->] (Z_4) edge [bend left = 30]  (S);
			\path [->] (X_2) edge (Y);

			\path [<->,dashed] (X_1) edge [bend left = 30] (Z_4);
			\path [<->,dashed] (X_2) edge [bend right = 30] (Z_2);
			\path [<->,dashed] (Y) edge [bend left = 30] (Z_2);
			\path [<->,dashed] (Y) edge [bend right = 30] (Z_4);
			
			\path [<->,dashed] (X_1) edge [bend left = 30] (X_2);
			\path [<->,dashed] (X_1) edge [bend left = 30] (Y);			
			\path [<->,dashed] (X_2) edge [bend right = 30] (Y);
			\path [<->,dashed] (Z_1) edge [bend left = 30] (Z_3);
			\path [->] (X_1) edge (S);
			\path [->] (X_2) edge (S);
			\path [->] (Z_1) edge (S);
			\path [->] (Z_3) edge (S);			
			\end{tikzpicture}
			\caption{Extended graph with four additional bidirected arcs and four additional arrows to $S$ \label{fig:correa3ab}}
			\end{subfigure}
\caption{Example 2: Two graphs where $P(Y \cond \doo(X_1,X_2))$ is identifiable from $P(Y, Z_1,Z_2,Z_3,Z_4 \cond \doo(X_1,X_2),S)$, $P(Z_1,Z_2)$ and $P(Z_3,Z_4)$. \label{fig:correa3a}} 
\end{figure}
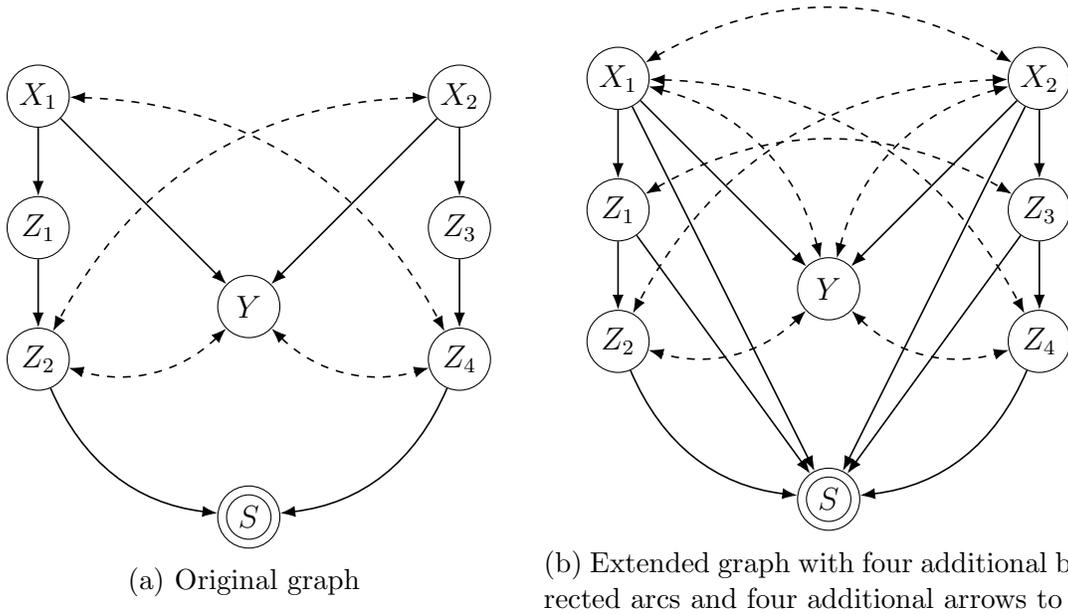

Figure~\ref{fig:correa3aa} presents another example by \citet{correa2019adjustment}.  Here the causal effect of interest is the joint effect of $X_1$ and $X_2$ on $Y$. An RCT provides information on this effect but suffers from selection bias $S$ that depends on $Z_2$ and $Z_4$. The RCT is symbolically written as $P(Y, Z_1,Z_2,Z_3,Z_4 \cond \doo(X_1,X_2),S)$. An observational data source is free from selection bias and provides information on $P(Z_1,Z_2,Z_3,Z_4)$. As it is noted in \citep{correa2019adjustment},  no subset of $\{Z_1,Z_2,Z_3,Z_4\}$ works as an adjustment set in this problem.

The application of do-calculus leads to the identifying formula
\begin{equation} \label{eq:ex2a}
P(Y \cond \doo(X_1,X_2)) =
\sum_{Z_2,Z_4}P(Z_2 \cond Z_1)P(Z_4 \cond Z_3)P(Y \cond \doo(X_1,X_2),Z_1,Z_2,Z_3,Z_4,S). 
\end{equation}
Figure~\ref{fig:Rcode} presents an R code that can be used to obtain this solution with the help of \texttt{do-search}. The outmost summation in equation~\eref{eq:ex2a} goes over variables $Z_2$ and $Z_4$. However, instead of joint distribution $P(Z_2,Z_4)$ that would be present in an adjustment, the formula contains the terms 
$P(Z_2 \cond Z_1)$ and  $P(Z_4 \cond Z_3)$. Here both $Z_1$ and $Z_3$ are trapdoor variables because they are functionally independent from the identifying functional and the causal effect is not identifiable if there are removed by a latent projection. Instead of the joint distribution $P(Z_1,Z_2,Z_3,Z_4)$ of four variables, pairwise distributions $P(Z_1,Z_2)$ and $P(Z_3,Z_4)$ are sufficient for identification.

\begin{figure}
\begin{center}
\begin{minipage}{17.3cm}
\begin{Verbatim}[fontsize=\small, frame=single, framesep=2mm, baselinestretch=0.9]
library(dosearch)

graph <- "
 X1 -> Z1
 X1 -> Y
 Z1 -> Z2
 Z2 -> S
 X2 -> Z3
 X2 -> Y
 Z3 -> Z4
 Z4 -> S
 X1 <-> Z4
 X2 <-> Z2
 Z2 <-> Y
 Z4 <-> Y
"

data <- "
    P(Y,Z1,Z2,Z3,Z4 | do(X1,X2),S)
    P(Z1,Z2)
    P(Z3,Z4)
"

query <- "P(Y|do(X1,X2))"

dosearch(data, query, graph, control = list(heuristic = FALSE)) 

-------------------------------------- Output --------------------------------------

"\sum_{Z2,Z4}\left(p(Z4|Z3)\left(p(Z2|Z1)p(Y|do(X1,X2),Z1,Z2,Z3,Z4,S)\right)\right)"
\end{Verbatim}
\end{minipage}
\end{center}
\caption{Example R code on the use \texttt{do-search} to determine identifiability of $P(Y \cond \doo(X_1,X_2))$ from inputs $P(Y,Z_1,Z_2,Z_3,Z_4  \cond \doo(X_1,X_2))$, $P(Z_1,Z_2)$  and $P(Z_3,Z_4)$ under the assumptions encoded in the graph of Figure \ref{fig:correa3aa}. 
}
\label{fig:Rcode}
\end{figure}

The causal effect $P(Y \cond \doo(X_1,X_2))$ can be also identified with purely observational data sources. Assume that we have information the distributions and $P(Z_2,Z_4)$ and $P(Y, Z_1,Z_2,Z_3,Z_4,X_1,X_2 \cond S)$. Using do-calculus we obtain 
\begin{equation} \label{eq:ex2aobs}
P(Y \cond \doo(X_1,X_2)) =
\frac{\sum_{Z_1,Z_2,Z_3,Z_4}P(Z_2,Z_4)P(Y,Z_1,Z_3,X_1,X_2 \cond Z_2,Z_4,S)}{\sum_{Y} \sum_{Z_1,Z_2,Z_3,Z_4}P(Z_2,Z_4)P(Y,Z_1,Z_3,X_1,X_2 \cond Z_2,Z_4,S)}. 
\end{equation}
Here the identifying functional takes a form of a fraction that represents a conditional distribution. After marginalizations, the numerator depends on $Y$, $X_1$ and $X_2$ while the denominator depends on $X_1$ and $X_2$. It can be confirmed with do-calculus that equation~\eref{eq:ex2aobs} indeed represents conditional distribution $P(Y \cond X_1, X_2)$. It is noteworthy that the joint distribution of only two variables, $Z_2$ and $Z_4$, is sufficient to recover from the selection bias.

Figure~\ref{fig:correa3ab} shows an extended version of graph of Figure~\ref{fig:correa3aa}. Despite the increased complexity of the graph, the identifying formula of equation~\eref{eq:ex2a} remains valid. However, $P(Y \cond \doo(X_1,X_2))$  is not identifiable anymore from observational data sources alone because there is an unobserved confounder between $X_1$ and $Y$ and between $X_2$ and $Y$.

\section{Simulations} \label{sec:numerical}

The simulation experiment is related to the graph of Figure~\ref{fig:ex1a} and uses an estimator based on equation~\eref{eq:ex1a} to estimate the average causal effect of $X$ on $Y$. We are interested in the accuracy of estimation as a function of sample sizes of the RCT and the survey. The data are generated according the following model:
\begin{align*}
U_{Z_1},U_{Z_1},U_{Z_3},U_{X},U_{Y},U_{X,Z_3},U_{Y,Z_3},U_S & \sim \textrm{N}(0,1), \\
Z_1 & = I(U_{Z_1} < 0), \\
X & = I(U_{X} < Z_1 + U_{X,Z_3}), \\
Z_2 & = I(U_{Z_2} < X), \\
Z_3 & = I(U_{Z_3} < Z_1 + Z_2 + U_{X,Z_3} + U_{Y,Z_3}), \\
Y & = I(U_{Y} < -1 + 2Z_1 + X + U_{Y,Z_3}), \\
S & = I(U_S < Z_3),
\end{align*}
where $I()$ stands for an indicator function. In other words, all unobserved variables are Gaussian and all observed variables are binary. In the domain $T$ where the RCT is conducted, we have 
\begin{equation*}
Z_1 \cond T  = I(U_{Z_1} < 1),
\end{equation*}
which changes the distribution of the background variable $Z_1$.
The simulation is carried out in R \citep{R} and package \texttt{R6causal} \citep{R6causal} is used to implement the data generation from the structural causal model.  

The estimator used is a plug-in estimator
\begin{equation} \label{eq:ex1aest}
\hat{P}(Y \cond \doo(X)) = \sum_{Z_1,Z_3}\left(\hat{P}(Y \cond \doo(X),Z_1,Z_3,Z_2,T,S)\sum_{X}\left(\hat{P}(Z_1,X)\hat{P}(Z_3 \cond Z_1,X,Z_2)\right)\right), 
\end{equation}
where $\hat{P}(Y \cond \doo(X),Z_1,Z_3,Z_2,T,S)$ is estimated from a subset of the RCT where either $Z_2=0$ or $Z_2=1$, $\hat{P}(Z_1,X)$ is estimated from the full survey, and $\hat{P}(Z_3 \cond Z_1,X,Z_2) $ is estimated from a subset of the survey where either $Z_2=0$ or $Z_2=1$ following the choice made for the RCT.  All probabilities are estimated as empirical proportions in the data.

The results of the simulation are reported in Table~\ref{tab:simresults}. The estimate $\hat{P}(Y=1 \cond \doo(X=1)) \approx 0.711$ seem to be biased for small sample sizes.  The choice $Z_2=1$ leads to smaller bias and smaller root mean square errors (RMSE) than $Z_2=0$ because in both the RCT and the survey the value $Z_2=1$ is more common than $Z_2=0$. More precisely, $\hat{P}(Z_2=1 \cond \doo(X=1)) = \hat{P}(Z_2=1 \cond X=1) \approx 0.841 $ and $\hat{P}(Z_2=1) \approx 0.715$.
 Starting from the sample size 400 for the RCT, the estimates appear to be practically unbiased for $Z_2=1$.

As expected, the RMSEs decrease as the number of observations increase in the RCT and in the survey. In this example, the sample size of the RCT seems to have larger impact than the sample size of the survey.  

\begin{table}
\caption{\label{tab:simresults} The bias and the root mean square error (RMSE) of the estimates of   $P(Y=1 \cond \doo(X=1))$ obtained using equation~\eref{eq:ex1aest} with choices $Z_2=1$ and $Z_2=0$. The data are generated according to the graph of Figure~\ref{fig:ex1a}. The number of simulation runs is 20000 for each combination of the sample sizes of the RCT and the survey.
}
\centering
\begin{tabular}{rrrrrr}
\multicolumn{2}{c}{Sample size} & \multicolumn{2}{c}{Bias} & \multicolumn{2}{c}{RMSE} \\
RCT & Survey & $Z_2=1$ & $Z_2=0$ & $Z_2=1$ & $Z_2=0$ \\
\hline
100  &  50  &  -0.009  &  -0.070  &  0.092  &  0.173 \\ 
 100  &  100  &  -0.009  &  -0.064  &  0.086  &  0.165 \\ 
 100  &  200  &  -0.009  &  -0.063  &  0.083  &  0.159 \\ 
 100  &  1000  &  -0.008  &  -0.062  &  0.082  &  0.154 \\ 
 100  &  10000  &  -0.009  &  -0.063  &  0.081  &  0.155 \\ 
 200  &  50  &  -0.001  &  -0.043  &  0.066  &  0.157 \\ 
 200  &  100  &  -0.001  &  -0.033  &  0.060  &  0.148 \\ 
 200  &  200  &  0.000  &  -0.031  &  0.058  &  0.144 \\ 
 200  &  1000  &  -0.001  &  -0.032  &  0.055  &  0.140 \\ 
 200  &  10000  &  -0.001  &  -0.032  &  0.054  &  0.139 \\ 
 400  &  50  &  -0.000  &  -0.022  &  0.052  &  0.122 \\ 
 400  &  100  &  0.000  &  -0.008  &  0.045  &  0.113 \\ 
 400  &  200  &  0.000  &  -0.007  &  0.041  &  0.107 \\ 
 400  &  1000  &  -0.000  &  -0.007  &  0.038  &  0.104 \\ 
 400  &  10000  &  0.000  &  -0.006  &  0.037  &  0.103 \\ 
 1000  &  50  &  -0.000  &  -0.015  &  0.042  &  0.085 \\ 
 1000  &  100  &  -0.000  &  -0.001  &  0.034  &  0.070 \\ 
 1000  &  200  &  -0.000  &  -0.000  &  0.029  &  0.064 \\ 
 1000  &  1000  &  -0.000  &  -0.000  &  0.024  &  0.060 \\ 
 1000  &  10000  &  0.000  &  0.000  &  0.023  &  0.059 \\
\end{tabular}
\end{table}

\section{Conclusions} \label{sec:discussion}
We have considered the generalization of experimental findings between populations and demonstrated that in settings where  no adjustments are applicable it is sometimes possible to apply do-calculus to derive more complicated identification strategies. 
Thanks to the flexibility of do-calculus, we are able to handle a large variety of problems with multiple data sources and intricate selection mechanisms. The difficulties of manual application of do-calculus can be avoided by using an algorithmic implementation \citep{tikka2021dosearch,dosearchR}. 

We also demonstrated that the identifying functional for the generalization of experimental findings may include trapdoor variables. The trapdoor variables in this paper are children of an intervened variable and on the path from the intervened variable to the selection variable while trapdoor variables presented in the literature have been ancestors of the intervened variable.  The values of trapdoor variables need to be fixed in estimation and this choice may have a considerable impact on the bias and accuracy of estimates when the sample sizes are small as it was demonstrated in the simulation. Whether an advanced estimation technique can reduce bias remains a question for future research. 

Besides the current setting with an RCT and some observational data sources, other settings could be considered as well. For instance, we could have data from a chain of experiments \citep{karvanen2020dosearch} or from a case-control study \citep{tikka2021dosearch}. Even in these settings, do-calculus is a powerful tool for synthesizing information from several data sources. 

\section{Acknowledgement}
CSC – IT Center for Science, Finland, is acknowledged for computational resources.

\bibliographystyle{apalike}
\bibliography{references}

\end{document}